\documentclass[%
 aip,
 amsmath,amssymb,
 reprint,%
reprint]{revtex4-1}

\usepackage{float}
\usepackage{graphicx}
\usepackage{dcolumn}
\usepackage{bm}
\usepackage[utf8]{inputenc}
\usepackage[T1]{fontenc}
\usepackage{mathptmx}
\usepackage{etoolbox}

\makeatletter
\def\@email#1#2{%
 \endgroup
 \patchcmd{\titleblock@produce}
  {\frontmatter@RRAPformat}
  {\frontmatter@RRAPformat{\produce@RRAP{*#1\href{mailto:#2}{#2}}}\frontmatter@RRAPformat}
  {}{}
}%
\makeatother
\begin{document}

\preprint{AIP/123-QED}

\title{On the practical applicability of modern DFT functionals for chemical computations. Case study of DM21 applicability for geometry optimization.}
\author{Kirill Kulaev}

\affiliation{ 
Skolkovo Institute of Science and Technology, Applied AI Center, Moscow, 121205, Russian Federation
}
\author{Alexander Ryabov}

\affiliation{ 
Skolkovo Institute of Science and Technology, Applied AI Center, Moscow, 121205, Russian Federation
}

\author{Michael Medvedev}

\affiliation{%
Zelinsky Institute of Organic Chemistry of Russian Academy of Sciences, Moscow, 119991, Russian Federation.
}

\author{Evgeny Burnaev}
\affiliation{ 
Skolkovo Institute of Science and Technology, Applied AI Center, Moscow, 121205, Russian Federation
}
\affiliation{ 
Autonomous Non-Profit Organization Artificial Intelligence Research Institute (AIRI), Moscow, 121170, Russian Federation
}

\author{Vladimir Vanovskiy}
\affiliation{ 
Skolkovo Institute of Science and
Technology, Applied AI Center, Moscow, 121205, Russian Federation
}

\date{\today}

\begin{abstract}
Density functional theory (DFT) is probably the most promising approach for quantum chemistry calculations considering its good balance between calculations precision and speed. In recent years, several neural network-based functionals have been developed for exchange-correlation energy approximation in DFT, DM21 developed by Google Deepmind being the most notable between them. This study focuses on evaluating the efficiency of DM21 functional in predicting molecular geometries, with a focus on the influence of oscillatory behavior in neural network exchange-correlation functionals.  We implemented geometry optimization in PySCF for the DM21 functional in geometry optimization problem, compared its performance with traditional functionals, and tested it on various benchmarks. Our findings reveal both the potential and the current challenges of using neural network functionals for geometry optimization in DFT. We propose a solution extending the practical applicability of such functionals and allowing to model new substances with their help.
\end{abstract}

\maketitle

%

\section{\label{sec:level1}Introduction}

Density Functional Theory (DFT) is a fundamental computational method widely used in chemistry and physics to investigate the electronic and nuclear structures of many-body systems, including atoms, molecules, and condensed phases. Its popularity comes from a good balance between computational efficiency and accuracy. More accurate computational methods, such as CCSD \cite{vcivzek1966correlation} and Quantum Monte Carlo \cite{mcmillan1965ground, metropolis1953equation}, can provide higher precision but are computationally intensive and scale poorly with system size. A central component of DFT is the exchange-correlation (XC) component of total energy, which encapsulates the complex many-body interactions among electrons. The exact form of the XC functional is unknown, and developing accurate approximations for it is essential because it determines the overall accuracy of DFT calculations. Traditionally, XC functionals have been constructed using analytical expressions based on approximations such as the Local Density Approximation (LDA) \cite{vosko1980accurate, perdew1981self, perdew1992accurate}, the Generalized Gradient Approximation (GGA) \cite{perdew1986accurate, perdew1992atoms, perdew1996generalized}, and Meta-GGA \cite{perdew1999accurate, tao2003climbing}. These functionals involve tuning multiple parameters within chosen mathematical forms to fit experimental data or results from more accurate calculations. Although successful in many applications, this approach struggles from limited flexibility and can make it challenging to incorporate new physical specificities from advanced computational methods like quantum Monte Carlo or post-Hartree-Fock calculations.

In fact, exchange-correlation functionals serve as \textit{surrogate models}— approximate mathematical models used to calculate the effects of electron-electron interactions within a system of formally non-interacting particles in DFT. Neural networks, being universal approximators \cite{Cybenko1992}, are naturally used to build surrogate models, so neural networks have become state-of-the-art approach in various fields: atomic potentials \cite{https://doi.org/10.48550/arxiv.2302.14231, https://doi.org/10.24433/CO.2788051.v1}, protein folding \cite{Jumper2021}, property prediction \cite{Sharma2023}. 
And also it offers the potential to achieve new levels of accuracy in DFT calculations considering the flexibility of neural network approximators, with their ability to handle a large number of parameters. In recent years, neural network-based functionals have seen significant development. These efforts include creation of completely new neural network based functionals \cite{Kirkpatrick2021-tg, Nagai2020, Dick2020-aq, Zhou2019} as well as studies aimed to interpolation or improving of existing functionals with the focus to analysis of physical conditions and new architectures \cite{Ryabov2020, Ryabov2022, Zheng2004}. Furthermore, some models incorporate physical asymptotic constraints to improve functional accuracy \cite{Nagai2022}. However, despite the high accuracy obtained in energy calculation, neural network functionals are not as commonly used in practical research as analytical functionals. The official documentation for quantum chemical packages like PySCF \cite{Sun2020}, Octopus \cite{tancogne2020octopus}, ORCA \cite{neese2020orca}, and others lacks comprehensive information about such functionals, and there is no full integration of such functionals by the developers of these packages, this can be attributed to several factors. Firstly, neural networks are often characterized by limitations such as low interpretability and vulnerability to adversarial attacks. Additionally, prior to the introduction of DM21, neural network functionals did not demonstrate superior performance compared to traditional analytical functionals. Secondly, there has been limited investigation into the application of these functionals in practical contexts, including geometric optimization, considering their oscillatory behaviour, computational cost,  impact of basis set selection, and the extrapolation capabilities beyond the training dataset. 
\ 

\par A neural network (NN) XC functional typically takes electron density descriptors as input and outputs the XC energy density. The total energy of the molecule is then obtained by integrating XC energy density over the grid and adding Coulomb and kinetic energies. The optimization of the neural network weights is achieved by minimizing the error between the predicted total energy and the reference total energy values. In practice, the energy difference between the systems is often used as a reference due to its physical and chemical significance.
\par However, the neural network approach to constructing XC functionals may encounter several issues. One of the main challenges is that neural network predictions can exhibit non-smooth behavior, particularly when calculating the derivatives of the exchange-correlation energy with respect to input variables such as electron density and its gradients.  This non-smoothness can lead to oscillations in the gradient, affecting the overall precision of the ground state energy obtained by main optimization cycle of DFT calculation, also called self-consistent field (SCF) cycle \cite{kohn1965self}. Consequently, such wiggle behavior on local scales can be noticeable in the calculation of the gradients of the exchange-correlation potential \(v_{xc}\) and the energy, which are crucial for optimization of geometry. 

\par  It is known that neural network functionals may not generalize well to systems containing elements or configurations not represented in the training data \cite{D4CP00878B}. Most datasets used for training focus on the energies of systems in stable geometries, meaning the extrapolation of these functionals to broader regions of the potential energy surface (PES) has not been systematically tested.  As a result, the reliability of these functionals in practical scenarios with geometry optimization remains an open question. To date, neural network functionals have primarily been tested for predicting reaction energies, calculated as the difference between several already relaxed geometries. Obtaining these relaxed geometries is a more complex problem than just estimating the energy. No quantitative evaluations of the accuracy of neural network functionals have been reported in geometric optimization task. 
\par In this work, we aim to systematically evaluate the performance of selected neural network functional DM21 \cite{Kirkpatrick2021-tg} in geometry optimization tasks. 
We investigate oscillatory behaviour and the resulting errors in the gradients on the optimization cycle results.
By conducting a quantitative analysis of the accuracy of neural network functional in geometry optimization, this study seeks to fill a crucial gap in the current body of research  and to build a bridge to its practical applications.

\section{\label{sec:level1}Methods}
\subsection{\label{sec:level2}Density functional theory}

Density functional theory is a quantum mechanical method widely used to investigate the electronic structure of many-body systems, including atoms, molecules, and solids. The fundamental principle of DFT is that all ground-state properties of a system of interacting electrons can be determined from the electron density \(\rho(\mathbf{r})\), which depends only on three spatial coordinates, rather than the many-electron wavefunction \(\Psi(\mathbf{r}_1, \mathbf{r}_2, \dots, \mathbf{r}_N)\), which depends on the coordinates of all \(N\) electrons and is computationally intractable for large systems.

\par The foundation of DFT was established by the Hohenberg-Kohn theorems \cite{Hohenberg_1964}. The first theorem states that the ground-state electron density uniquely determines the external potential (and thus the Hamiltonian and all properties of the system). The second theorem introduces a variational principle: the ground-state energy is a functional of the electron density, and the correct ground-state density minimizes this energy functional.

\par In practice, DFT calculations are performed using the Kohn-Sham formalism \cite{Hohenberg_1964, kohn1965self}, which maps the complex interacting electron system onto a fictitious system of non-interacting electrons moving in an effective potential. This approach simplifies the problem while preserving the exact ground-state electron density. The Kohn-Sham equations are:

\begin{equation} \label{eq:KS}
\left[ -\frac{\hbar^2}{2m} \nabla^2 + v_{\text{ext}}(\mathbf{r}) + v_{\text{H}}(\mathbf{r}) + v_{\text{xc}}(\mathbf{r}) \right] \phi_i(\mathbf{r}) = \varepsilon_i \phi_i(\mathbf{r}),
\end{equation}

where:
\begin{itemize}
    \item \(\phi_i(\mathbf{r})\) are the Kohn-Sham orbitals,
    \item \(\varepsilon_i\) are the corresponding orbital energies,
    \item \(v_{\text{ext}}(\mathbf{r})\) is the external potential due to the nuclei and any applied fields,
    \item \(v_{\text{H}}(\mathbf{r})\) is the Hartree potential representing the classical electrostatic (Coulomb) interaction between electrons,
    \item \(v_{\text{xc}}(\mathbf{r})\) is the exchange-correlation potential, accounting for quantum mechanical effects of exchange and correlation between electrons.
\end{itemize}

\par The exchange-correlation potential \(v_{\text{xc}}(\mathbf{r})\) is defined as the functional derivative of the exchange-correlation energy \(E_{\text{xc}}[\rho]\) with respect to the electron density:

\begin{equation}
\label{eq2}
\begin{split}
v_{\text{xc}}(\mathbf{r}) = \frac{\delta E_{\text{xc}}[\rho]}{\delta \rho(\mathbf{r})},
\end{split}
\end{equation}

Total exchange-correlation energy can be written as (also called exchange-correlation functional):

\begin{equation}
\begin{split}
E_{xc} = \int_{}^{}f_{xc}[\rho(\mathbf{r}), \nabla \rho(\mathbf{r}), ...]d\mathbf{r},
\end{split}
\label{total_Exc}
\end{equation}

Then the total potential energy of system $E_{tot}$ is calculated from $E_{ext}$  - the energy due to the external potential, Coulumb energy $E_{J}$ and exchange-correlation energy $E_{xc}$.

\begin{equation}
\begin{split}
E_{tot} = E_{ext} + E_{J} + E_{xc},
\end{split}
\label{E_tot}
\end{equation}

Challenge in DFT lies in approximating \(E_{\text{xc}}[\rho]\) accurately since its exact form is unknown. The simplest approximation is the Local Density Approximation (LDA) \cite{vosko1980accurate, perdew1981self, perdew1992accurate}, which assumes that the exchange-correlation energy at each point depends only on the local electron density. While LDA provides reasonable accuracy for simple systems with slowly varying electron densities, it struggles with more complex systems where nonlocal effects become important.  To address nonlocal effects, additional characteristics of the electronic structure, such as the electron density gradient \(\nabla\rho(\mathbf{r})\) in the Generalized Gradient Approximation (GGA) \cite{perdew1986accurate, perdew1992atoms, perdew1996generalized} and the kinetic energy density \(\tau(r) = \frac{h^{2}}{2m}\sum_{i}^{}\left| \nabla\phi_{i}(\mathbf{r}) \right|^{2} \) in Meta-Generalized Gradient Approximation (Meta-GGA) \cite{perdew1999accurate, tao2003climbing} are incorporated into the exchange-correlation functional. Moreover, introduction of orbital-dependent energies, such as in hybrid functionals \cite{becke98density} significantly increases the level of approximation. Neural network functionals functionals can belong to different levels of approximation (LDA, GGA etc.).

\subsection{\label{sec:level2}DM21 Functional Overview}

DM21 is one of modern neural network XC functionals. It achieved qualitatively new results for systems with strong correlation and artificial charge delocalization. 
According to the detailed energy benchmark conducted in the original paper \cite{Kirkpatrick2021-tg}, DM21 significantly outperformed classical functionals in energy calculation of main-group molecules and atoms, the mean absolute error (MAE) in energy values on the GMTKN55 benchmark \cite{Goerigk2017} for DM21 was 1.5 kcal/mol, compared to 3.6 kcal/mol using analytical functional SCAN \cite{Sun2015}.
\par
DM21 depends on the local Hartree-Fock exchange energies \(e^{HF}\) and it's range-separated version \(e^{\omega HF}\). This approach positions DM21 as a Local Hybrid range-separated Functional (LHF). Instead of mixing the global Hartree-Fock exchange energy with the integrated value of the semi-local XC functional, LHFs \cite{Maier2018} mix the value of the Hartree-Fock exchange energy at point \(\mathbf{r}\) with the output of the XC functional in this point. The expression for the local Hartree-Fock exchange is expressed by the formula, where \(D_{pq} = \sum_{a}^{occupied}c_{p a} c_{q a}
\) is the density matrix, \(c_{a}\)  is the orbital coefficient and \(\phi_{a}\) represents the basis orbitals:

\begin{equation} \label{eq10}
\begin{split}
e_{x}^{LDA}(\mathbf{r}) = -2 \pi[(3/4\pi)(\rho^{\uparrow}(\mathbf{r})+\rho^{\downarrow}(\mathbf{r}))]^{4/3},
\end{split}
\end{equation}


\begin{multline}\label{eq5}
e^{\omega HF}_{\sigma}(\mathbf{r}) = -\frac{1}{2}\sum_{pqrs}^{}D^{\sigma}_{pq}D^{\sigma}_{rs}\\\times\int_{}^{}\phi_{p, \sigma}(\mathbf{r})\phi_{s, \sigma}(r)\frac{erf(\omega\left| \mathbf{r}-\mathbf{r}_{0} \right|)}{\left| \mathbf{r}-\mathbf{r}_{0} \right|}\phi_{r, \sigma}(\mathbf{r}^{\prime})\phi_{q, \sigma}(\mathbf{r}^{\prime})d^{3}\mathbf{r},
\end{multline}

Local Hartree-Fock exchange energy $e^{HF}_{\sigma}(\mathbf{r})$ is calculated similarly to $e^{\omega HF}_{\sigma}(\mathbf{r})$ with the $\omega$ range-separation parameter is zero. The expression of the exchange-correlation functional of DM21 is as follows.
\begin{equation} \label{eq6}
\begin{split}
E^{MLP}_{xc}[\rho] = \int_{}^{}f_{\theta}(x(\mathbf{r}))\cdot \begin{bmatrix}
 e_{x}^{LDA}(\mathbf{r})   \\
 e^{HF}(\mathbf{r})    \\
 e^{\omega HF}(\mathbf{r})
\end{bmatrix}
d^{3}\mathbf{r},
\end{split}
\end{equation}
\
\par\(f_{\theta}(x(r))\) is a vector of 3 values which are outputs of neural network. The integrand is computed by scalar product of \(f_{\theta}(x(r))\) and the vector of 3 energies at $\mathbf{r}$: $e_{x}^{LDA}(\mathbf{r})$, \(e^{HF}(\mathbf{r})=e^{HF}_{\uparrow}(\mathbf{r})+e^{HF}_{\downarrow}(\mathbf{r})\) and \(e^{\omega HF}(\mathbf{r})=e^{\omega HF}_{\uparrow}(\mathbf{r})+e^{\omega HF}_{\downarrow}(\mathbf{r})\). Neural network takes as input 11 variables including electron density for each spin channel, norms of electron density gradient, kinetic energy density, and four features of local Hartree-fock exchange energy.

Resolution of identity (RI) \cite{Ren2012} is a technique that is claimed to reduce the scaling of DM21 with system size and number of basis orbitals. 
 Unfortunately, since the release of DM21, which showed an increase in computational speed using this feature, no RI implementation for DM21 has been released that can replicate author's speed test. For this reason, in the speed test we describe below, we did not use speedup techniques for all methods. 

 Despite the high accuracy of the DM21 functional in energy calculations, this does not necessarily guarantee high accuracy in geometric optimization. Such optimization relies on the gradients of the energy with respect to the atomic nuclei coordinates (forces). These gradients can be highly oscillatory, possibly because such functionals are trained primarily on optimal or near-optimal molecular geometries, and may not accurately capture the energy landscape far from equilibrium configurations, leading to unstable optimization. While calculating energies and various properties for a set of known geometries—such as those previously computed by others—can be valuable, it is often necessary to perform a geometry optimization process if a suitable geometry is not readily available. This step ensures that the structure corresponds to a local or global minimum on the potential energy surface, which is critical for accurate energy and property predictions.

\subsection{\label{sec:level2}Geometry optimization}
The goal of geometry optimization is to find the coordinates of atoms $\mathbf{R}^*$ corresponding to the minimum energy on potential energy surface for given atomic structure represented as atomic coordinates $\mathbf{R}$. Then the optimization problem can be written as follows:
\begin{equation} \label{eq:geomopt}
\mathbf{R}^* =  \operatorname*{argmin}_{\mathbf{R}}E_{tot}(\mathbf{R}),
\end{equation}
Given XC functional and basis set, it starts from the initial coordinates of atoms. The process of geometric optimization using DFT can be represented as the following iterative algorithm:

\begin{figure}[H]
    \centering
    \includegraphics[width=0.45\linewidth]{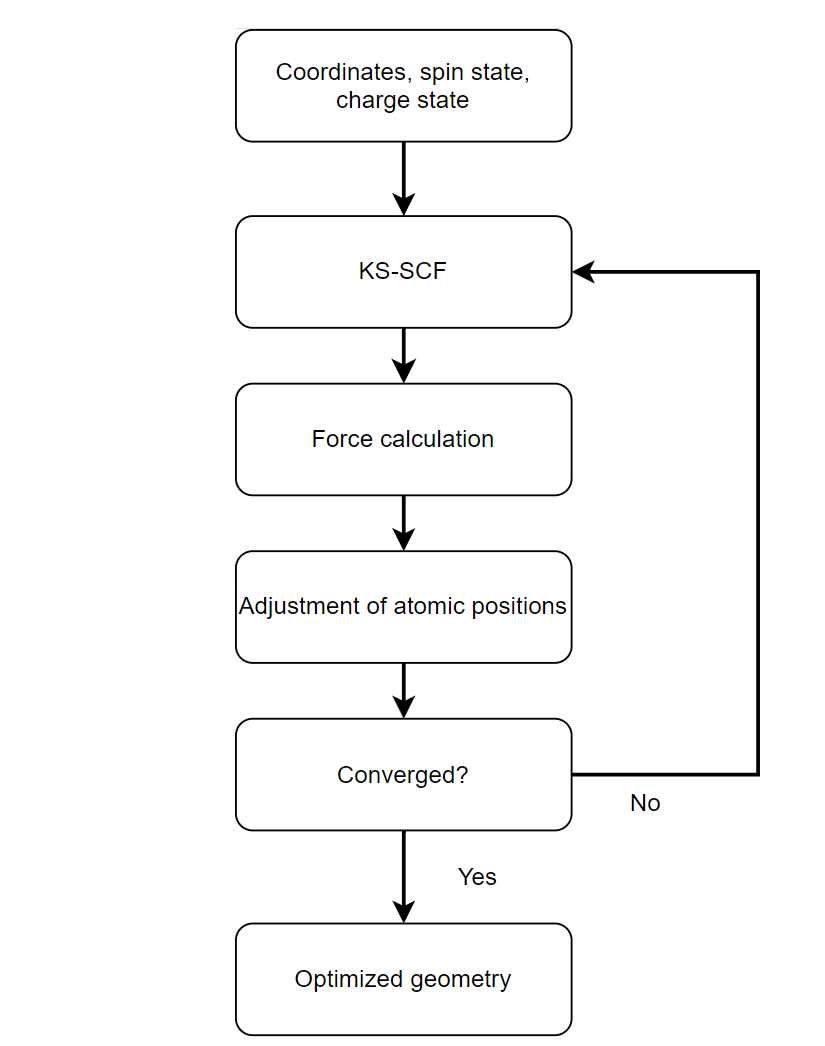}
    \caption{Schematic diagram of the DFT geometric optimisation algorithm }
    \label{fig:geomopt_scheme}
\end{figure}

The geometry optimisation Fig. \ref{fig:geomopt_scheme} starts by solving the Kohn-Sham equations using the SCF method. Then the forces are calculated (using analytical or numerical gradients) and the positions of the atoms are adjusted by the optimisation algorithm to minimise the energy. In the last step, convergence criteria are checked: changes in energy, forces or geometry adjustments.






\par In this work, the functionals PBE0\cite{Adamo1999}, SCAN\cite{Sun2015}, and DM21 were compared in predicting the geometries of molecules. For all systems in this study, we utilized Cartesian coordinate-based geometry optimization from the GeomeTRIC\cite{Wang2016-tr} library implemented in PySCF\cite{Sun2020}.

The GeomeTRIC library\cite{Wang2016-tr} introduces an approach to geometry optimization by employing a translation-rotation-internal coordinate (TRIC) system. This system explicitly includes the collective translations and rotations of molecules or molecular fragments. Translations are represented by the centroid position of the molecule, while rotations are parameterized using the exponential map of quaternions.

\subsection{\label{sec:level2}Nuclear gradients}

In each step of geometric optimization, the calculation of interatomic forces is essential. The force acting on an atom A is given by: 
\begin{equation} \label{eq:forces}
\mathbf{F}_A = -\nabla_{A} E(\mathbf{R}),
\end{equation}
\par  Previous studies have indicated that the calculation of energy derivatives with respect to atomic coordinates for local hybrid functionals is challenging \cite{Klawohn2016-jc}.  While there are semi-numerical approaches available \cite{Klawohn2016-jc}, there is currently no effective implementation of analytical gradients for local range-separated hybrid functionals, such as DM21, in the literature. To address this limitation, we have integrated numerical gradient calculations within the PySCF framework specifically for DM21.

\subsection{\label{sec:level2}Force calculation techniques}

\begin{figure*}
    \includegraphics[width=0.81\linewidth]{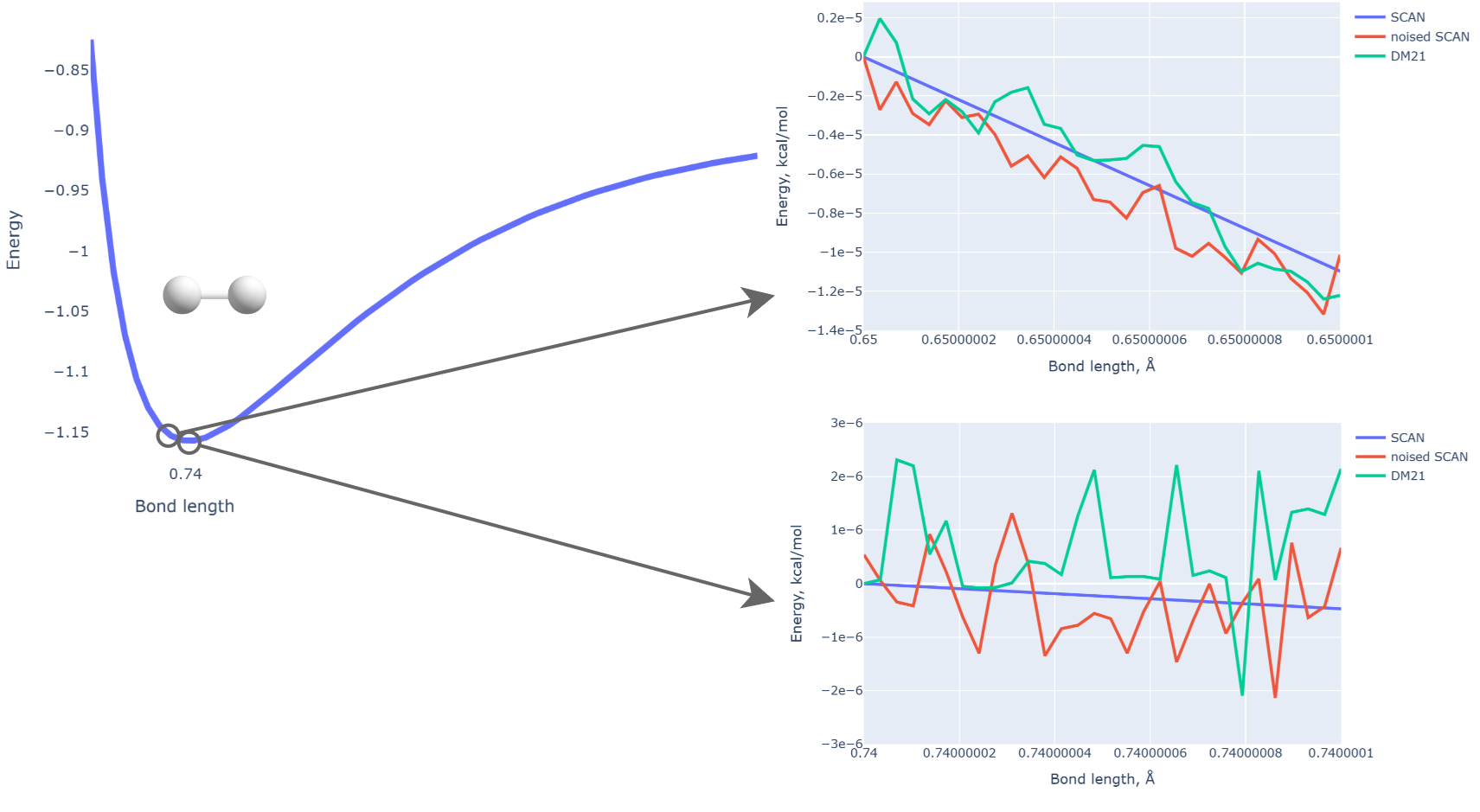}
    \caption{PES of H\(_{2}\) molecule with a shortened bond length (left) in minimum (right) calculated with DM21, SCAN, and SCAN with noise in \(e_{xc}\) and \(v_{xc}\). All the points were obtained from converged SCF single point calculations with energy tolerance of \(10^{-11} E_{h}\) (Hartree energy).}
    \label{fig:PES_noise}
\end{figure*}

The central finite difference method was used to calculate the interatomic forces. Force acting on the atom A in the finite difference approximation was calculated as:
\begin{equation} \label{eq:fd}
\mathbf{F}_A=-\frac{dE(\mathbf{R})}{d\mathbf{R}}\approx -\frac{E(\mathbf{R}+\Delta \mathbf{R})-E(\mathbf{R}-\Delta \mathbf{R})}{2\Delta \mathbf{R}},
\end{equation}
\par The central finite difference method has limited practicality due to the fact that the calculation of forces on one iteration of the geometry optimization requires \(6N+1\) separate SCF runs where \(N\) is the number of atoms. We accelerated numerical force calculations using the following technique: by making small increments to the atomic coordinates, we initiated the self-consistent energy calculation at the points \(\mathbf{R} \pm \Delta \mathbf{R}\) based on the electron density at \(\mathbf{R}\) . This approach allows us to reuse the density matrix from \(\mathbf{R}\)  in the self-consistent field calculations at \(\mathbf{R} \pm \Delta \mathbf{R}\). The numerical differentiation parameter $\Delta \mathbf{R}$ was chosen empirically, the selection of the optimal parameter for each system under study is intractable for practical applications (and this is a limitation of the method), for this reason we have checked the stability of the numerical forces for typical chemical bonds in molecules similar to the substances from the test sets under study and presented the results in Supplementary Materials.

\subsection{\label{sec:level2}Mathematical modeling of oscillations}\

Neural networks are known to suffer from non-smooth behavior of interpolation that produces the high-frequency oscillations in their gradients. The purpose of this subsection is to measure the level of oscillations in potential energy surface and their effect on the numerical and analytical nuclear gradients. Since DM21 does not have analytical nuclear gradients, we assumed that an analytical functional with added noise, the level of which is similar to the neural network one, would allow us to model the gradient behavior while preserving the possibility of both analytical and numerical estimates. To better understand and quantify the impact of these oscillations, we start with creating a mathematical model that captures its characteristics and effect on the potential energy surface (PES) in case of DM21 functional. Due to the fact that the magnitude of noise can vary with the elemental composition of substances \cite{D4CP00878B}, we chose the H$_2$ dissociation curve as a model system, when the noise in our model was calibrated on the SCF of H$_2$ and He$_2$ in order to then estimate the oscillations in the nuclear gradients on the H$_2$ system.


 Potential energy surface is a hypersurface that represents the total energy of a system as a function of atoms' positions (or related parameters). The total exchange-correlation energy \(E_{xc}\) is given by the integral over the electron density \(\rho(\mathbf{r})\) and the exchange-correlation energy per particle \(\epsilon_{xc}(\mathbf{r})\) \ref{total_Exc}. Within the scales of variations of input parameters used for the calculation of numerical derivatives to compute forces, \(\epsilon_{xc}\) and \(v_{xc}\) depend linearly on the inputs. We assume that the error in the inputs is modeled as Gaussian noise. For small variations in the input data \(\delta X\), the outputs \(\epsilon_{xc}\) and \(v_{xc}\) can be approximated linearly, and if the input data has Gaussian noise, the changes in the outputs will also linearly depend on this noise, thus following a normal distribution.  

 Therefore, this assumption allows us to create a tractable model for comparison. In order to model the noise of the DM21 neural network, we assumed Gaussian noise was added to the XC energy per particle and XC potential of SCAN functional.

\begin{equation}
\epsilon_{xc}' = \epsilon_{xc} + \eta_{\epsilon}, \quad \eta_{\epsilon} \sim \mathcal{N}(0, \sigma_{\epsilon}^2), \quad \sigma_{\epsilon} = 10^{-6.8},
\end{equation}

\begin{equation}
v_{xc}' = v_{xc} + \eta_{v}, \quad \eta_{v} \sim \mathcal{N}(0, \sigma_{v}^2), \quad \sigma_{v} = 10^{-6.2},
\end{equation}

The standard deviation of the noise was chosen to achieve two objectives: first, to ensure that the resulting energy oscillations matched the level of those calculated with DM21, and second, to align the number of SCF iterations with the number required for convergence with DM21. Note that for the H\(_{2}\) system in 6-31G basis set, the number of SCF iterations required for convergence with SCAN was 4, while with DM21, it was 16. The paper \cite{D4CP00878B} also highlighted the difficulty of achieving SCF convergence with DM21. Therefore, the increased number of SCF iterations with DM21 is attributed to the noise in \(e_{xc}\) and \(v_{xc}\).

Fig. \ref{fig:PES_noise} shows a detailed analysis of H\(_{2}\) molecule PES in the two regions, comparing the SCAN functional, SCAN with added noise, and the DM21 functional. The left side of the figure illustrates how the noise affects the PES before the equilibrium bond length, while the right side zooms in on the region closer to the minimum. The analysis demonstrates that the DM21-like noise leads to oscillations in the energy values, which are not present in the smooth PES generated by SCAN. We note that adding noise to the SCAN functional does not lead to significant deviations in forces.  Table. \ref{tab:tab1}. lists deviations in the forces with noisy SCAN calculated by numerical and analytical methods compared to unnoised analytical SCAN forces on the H\(_{2}\) dissociation curve. The pointed level of noise is exposed in incorrect calculations 
of the numerical gradient of energy at \(\Delta \mathbf{R}\) < 1e-7 Å, at a larger step of numerical differentiation the forces almost completely match those obtained in analytical form. 

\begin{table}
\caption{\label{tab:tab1}Deviations in nuclear gradient produced by noise in the \(e_{xc}\) and \(v_{xc}\).}
\begin{ruledtabular}
\begin{tabular}{cccccc}

Calculation Method & \(\Delta \mathbf{R}\), Å & MAE \\
\hline
Numerical & 1e-8 & 0.127 \\
Numerical & 1e-7 & 0.023 \\
Numerical & 1e-6 & 0.020 \\
Analytical & - & 2.5e-7 \\

\end{tabular}

\end{ruledtabular}
\end{table}

We thus estimated the contribution of DM21-like oscillations to nuclear gradients. According to the results, we can say that the oscillation contribution is minor, which further allows us to use the numerical force calculation method for geometric optimisation and to determine the efficiency of DM21 in the real molecular geometry optimisation problem. In the "Benchmark testing" subsection, we perform testing to analyze the impact of this noise on real geometry optimizations more thoroughly. This will provide a clearer understanding of how the DM21 functional compares with traditional functionals in practical applications.

\begin{table*}[ht]
\centering
\caption{\label{tab:tab2}Metrics of the tested XC functionals on benchmarks.}
\begin{ruledtabular}
\begin{tabular}{ccccc}
Benchmark & Basis set & DM21 MAE & PBE0 MAE & SCAN MAE \\
\hline
LMGB35 bond lengths (pm) & 6-31G(d,p) & 0.843 & 0.803 & 0.795 \\
LMGB35 bond lengths (pm) & def2-TZVP & 0.625 & 0.948 & 0.595 \\
LMGB35 energies (kcal/mol) & cc-pVQZ & 2.151 & 3.757 & 3.978 \\
\textit{ab-initio} geometries (pm) & 6-31G(d,p) & 1.906 & 1.957 & 1.987 \\
\textit{ab-initio} energies (kcal/mol) & cc-pVQZ & 0.677 & 0.598 & 0.297 \\
\end{tabular}
\end{ruledtabular}
\end{table*}

\section{\label{sec:level1}Results}

\subsection*{Speed tests}

\begin{figure}[H]
    \centering
    \includegraphics[width=0.85\linewidth]{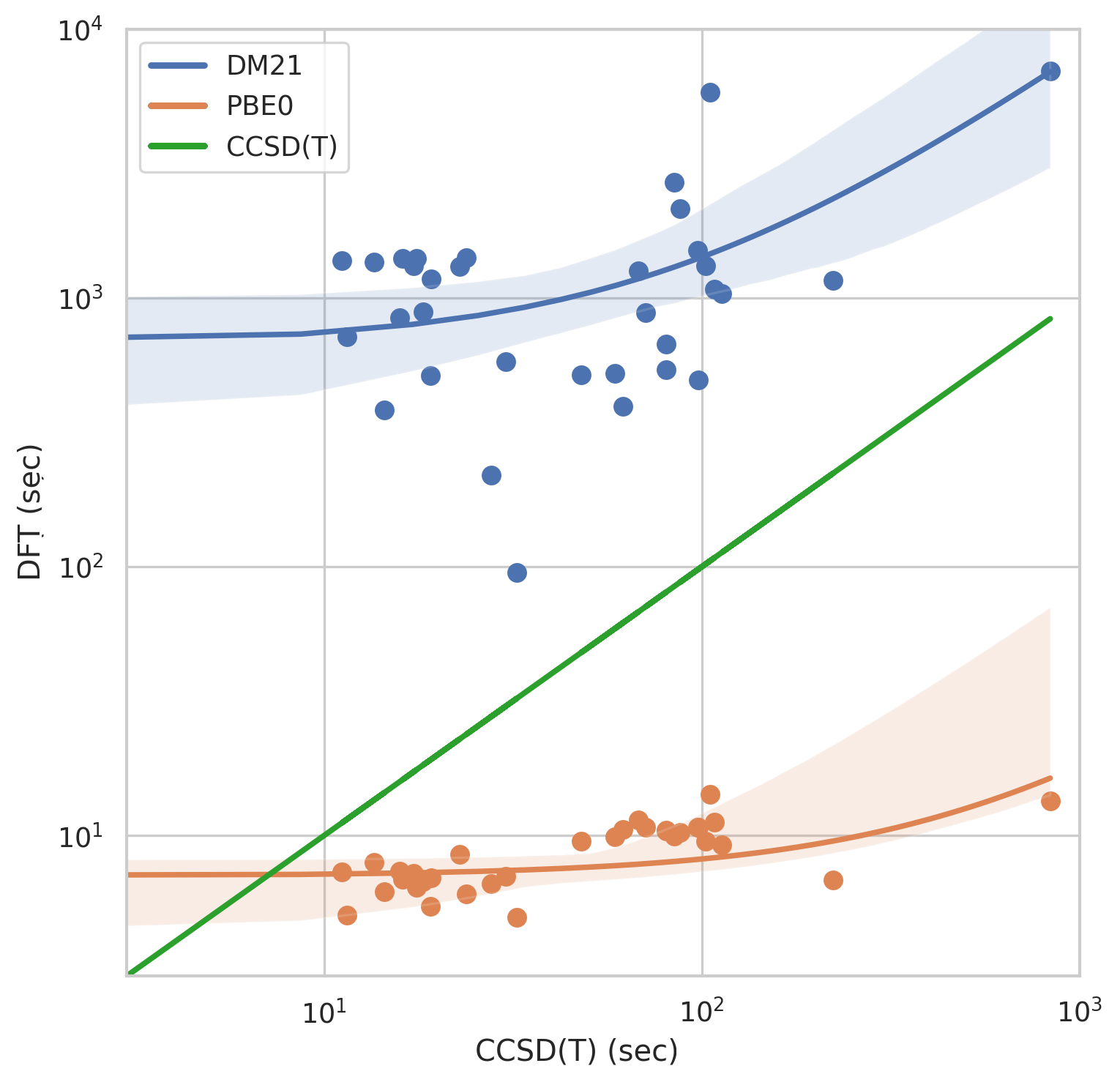}
    \caption{Time tests of PBE0, DM21 and CCSD(T). Horizontal axis shows the computation time of the above systems using CCSD(T) method, while vertical axis shows DFT computation time of these systems with DM21 and PBE0 functionals. Calculation have performed at Google Colab machine with 2 CPU cores of Intel(R) Xeon(R) CPU @ 2.20GHz and 13GB RAM.}
    \label{fig:time_tests}
\end{figure}

Geometry optimization is a combination of single-point computations. Therefore, we first determine how computationally expensive the SCF calculation with DM21 is compared to PBE0 and CCSD(T) (the main reference method for neural network training). For this purpose, we determined the calculation time of single SCF cycles for two-atom molecules of elements from H to Ca in neutral form and +1 cations using the cc-pVTZ basis set and energy tolerance of \(10^{-9} E_{h}\). The comparison of the calculation time using the mentioned methods is given in Fig. \ref{fig:time_tests}.

\begin{figure*}[ht]
\includegraphics[width=1\linewidth]{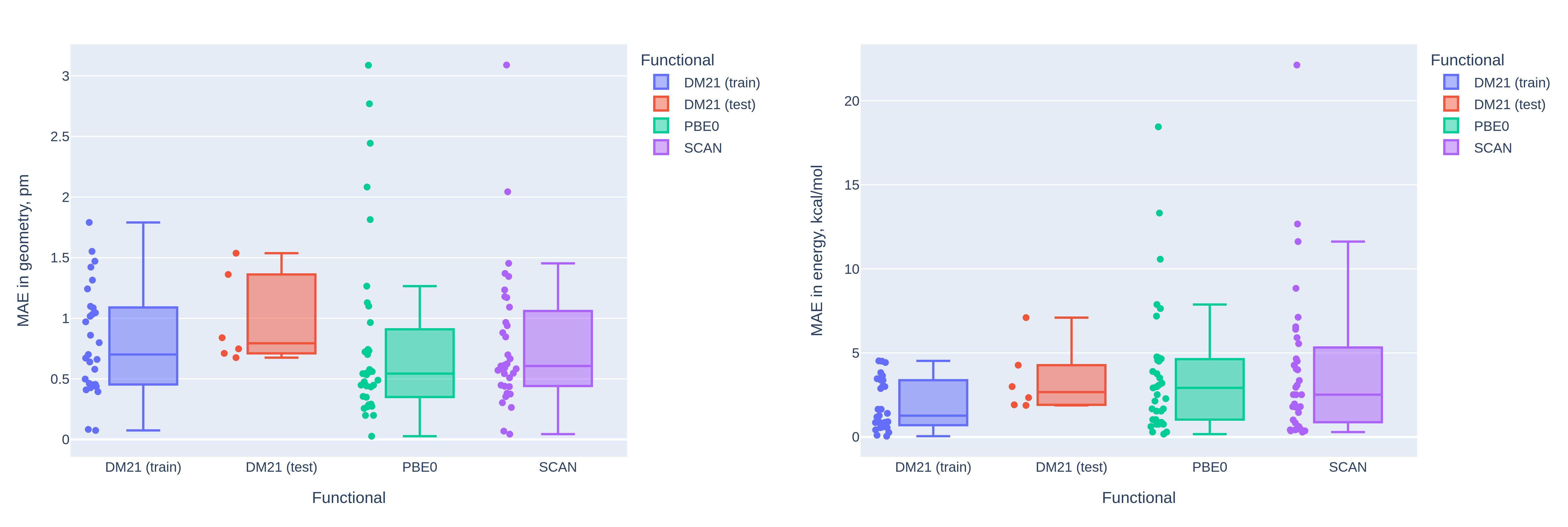}
\caption{Comparison of DM21 and analytical functionals on the LMGB35 benchmark}
\label{fig:lbgm35}
\end{figure*}

\begin{figure*}[ht]
    \centering
    \includegraphics[width=0.9\linewidth]{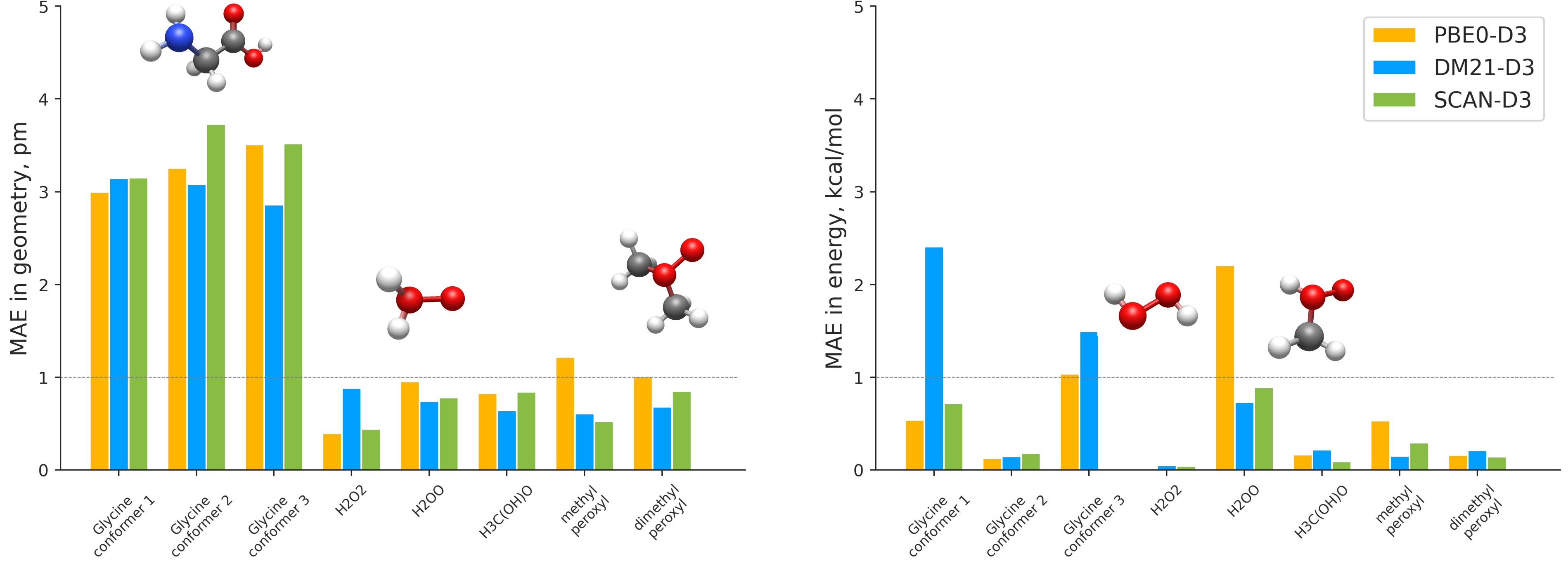}
    \caption{Benchmark of high-level optimized geometries}
    \label{fig:ab_initio_benchmark}
\end{figure*}

 Each point in Fig. \ref{fig:time_tests} shows the time taken to calculate this system using a labeled functional, thus points located above the green curve spent more calculation time than CCSD(T). It is worth noting that the DM21 implementation has a longer computation time, at least for small systems, than CCSD(T). This is directly related to: the need to compute two-electron integrals at each grid point, the difficulty of convergence of the SCF cycle due to noise. Thus, even if analytic gradients could be made for DM21, it would certainly speed up geometric optimization, but overall SCF DM21 is less efficient than CCSD(T) in terms of speed and accuracy (methodologically). 

\subsection{\label{sec:level2}Benchmark study}

\par LMGB35 \cite{Caldeweyher2019} is a benchmark for geometry optimization problem described in subsection "Geometry optimization". It contains covalent systems from the first and second rows of the periodic system compiled from experimental data, the reference values are represented by bond lengths. With the exception of systems such as BO, F\(_2^+\), N\(_2^+\), NF, NH\(^+\), O\(_2^+\), OH\(^+\), molecules from LMGB35 were represented in the DM21 training dataset (within the W4-17 subset). Calculations of DFT atomization energies were performed using the cc-pVQZ basis set, when its reference values were calculated using CCSD(T)/cc-pVQZ. Geometry optimizations were conducted with the 6-31G(d,p) and def2-TZVP basis sets.

Based on Fig. \ref{fig:lbgm35}, several conclusions can be drawn. Overall, the DM21 functional (blue bars) demonstrates competitive performance compared to the other functionals: PBE0 (yellow bars) and SCAN (green bars). The chart shows that the MAE in bond lengths after optimization is generally below 2 pm for most molecules across all functionals. There are specific molecules where DM21 shows lower MAE in bond lengths compared to PBE0 and SCAN, such as certain hydrocarbons (e.g., C\(_2\)H\(_4\), C\(_2\)H\(_2\)). However, there are also many cases where PBE0 or SCAN outperform DM21. Therefore, even though the molecules in the benchmark set are very similar to those in the training set of DM21, this similarity does not guarantee good geometry optimization capabilities. From the MAE values, one may observe that for the LMGB35 benchmark DM21 outperforms the analytical functionals in the energy calculation, but is not more accurate in the geometry optimization.

To assess accuracy beyond the training sample, DM21 was tested in geometry optimization of larger molecules that were not in the training set. The reference data were geometries optimized at the CCSD(T) level with a large basis set, as taken from the studies \cite{Olive2021, Gruber2023}. 

DM21 showed stable result in predicting geometries of glycine conformers (Fig.  \ref{fig:ab_initio_benchmark}), overall its accuracy on this benchmark was marginally higher than PBE0 and SCAN functional's accuracy. Despite the exclusion of these molecular systems from the training sample, generalizability of DM21 to this structures was confirmed.
Table. \ref{tab:tab2} summarizes the mean absolute error of each studied functional in the geometry and energy prediction tasks on the mentioned benchmarks.

\section{\label{sec:level1}CONCLUSION}
In our study, we performed the first analysis of neural network functional in geometry optimization tasks. Firstly, we conducted mathematical modeling of oscillations produced by NN and measured its impact to PES and gradients of NN. The results of modeling show that both analytical and numerical gradients can be used with NN XC functionals even with the presence of NN-specific oscillations. After that, we conducted several benchmark tests to assess practical applicability of DM21 functional. Firstly, we performed time tests of DM21 and compared its speed with PBE0 and CCSD(T). These tests showed that the speed of DM21 in single-point computations is strongly biased in favor of analytic functionals and even CCSD(T). Moreover, given the absence of analytic gradients for LHF, the geometric optimization problem for DM21 becomes practically limited. After that, we conducted comparison of DM21 and several analytical functionals in energy computations and geometry optimization. 

In summary, the analyses performed in this study have shown that, despite the significantly increased accuracy of energy calculation using the DM21 functional compared to the analytical one, the quality of equilibrium geometry prediction remains at nearly the same level as can be achieved using the analytical PBE0 and SCAN functionals. Additionally, the features of neural network functionals often do not allow for the analytical calculation of atomic forces, which significantly reduces their practical applicability. This may indicate the need to revise the approach to the construction and training of neural network XC functionals.

\par An effective way to increase the accuracy of exchange-correlation functionals, not only for the geometric optimization problem but also in general, involves informing the neural network about the curvature of the PES. For instance, when training the DM21 functional, such an approach was used in dataset construction by adding diatomic molecules with non-equilibrium bond lengths to the training sample. However, one can explicitly consider the PES curvature for functionals having an expression for the analytic derivative of energy at atomic coordinates (e.g., LDA, GGA, MGGA, hybrids, range-separated hybrids, and nonlocal correlation).  Incorporating into the loss function not only the error in energy but also the error in the derivatives with respect to atomic coordinates can adjust the functional to explicitly account for the local curvature of the PES. Consequently, experimental values of equilibrium geometries can serve as accurate references for the PES minima. Supposedly this approach could reduce overfitting and the amount of data necessary for training, thereby improving the generalizability of the functional, especially in complex cases for DFT.

\section*{\label{sec:level1}DATA AVAILABILITY STATEMENT}
The data that support the findings of this study are available from the corresponding author upon request.

\section*{\label{sec:level1}ACKNOWLEDGMENTS}

The authors are grateful to the support of the Russian Science Foundation (grant {\#}24-41-02035)

\section*{\label{sec:level1}AUTHOR DECLARATIONS}
\subsection*{\label{sec:level2}Conflict of Interest}
 The authors have no conflicts to disclose.

\section*{\label{sec:level1}REFERENCES}

\bibliography{references}

\end{document}